\documentstyle[pre,aps,epsfig]{revtex}

\begin{document}

\markboth{Joachim Krug}
{Coarsening of vortex ripples}

\title{COARSENING OF VORTEX RIPPLES IN SAND}

\author{\footnotesize JOACHIM KRUG}

\address{Fachbereich Physik, Universit\"at Essen\\
45117 Essen, Germany
}

\maketitle

\begin{abstract}
The coarsening of an array of vortex ripples prepared in an unstable
state is discussed within the framework of 
a simple mass transfer model 
first introduced by K.H. Andersen et al. 
[Phys. Rev. E {\bf 63}, 066308 (2001)]. Two scenarios for the 
selection of the final pattern 
are identified. When the initial state is homogeneous with uniform
random perturbations, a unique final state is reached which depends
only on the shape of the interaction function $f(\lambda)$. A
potential formulation of the dynamics suggests that the final wavelength
is determined by a Maxwell construction 
applied to $f(\lambda)$, but comparison
with numerical simulations shows that this yields only an upper bound.
In contrast, the evolution from a perfectly homogeneous state
with a localized perturbation proceeds through the propagation
of wavelength doubling fronts. 
The front speed can be predicted by standard marginal stability theory.
In this case the final wavelength depends on the initial wavelength
in a complicated manner which involves multiplication by factors
of 2 and rational ratios such as 4/3.

\end{abstract}

\vspace{1.cm} 

\small

\noindent
\emph{Uns \"uberf\"ullts. Wir ordnens. Es zerf\"allt.} \\
\emph{Wir ordnens wieder und zerfallen selbst.} \\
Rainer Maria Rilke\footnote{From the eighth  Duino elegy.}

\normalsize

\section{Introduction}

Vortex ripples are a familiar occurrence in coastal waters, where
the waves expose the sand surface at the sea bottom to an oscillatory flow.
The name reflects the important r\^{o}le of the separation vortices that
form on the lee side of the ripples in stabilizing the ripple slopes
\cite{Ayrton10,Bagnold46}.
In laboratory experiments, one observes the formation of a stable periodic
pattern with a ripple wavelength proportional to the amplitude $a$ of 
the water motion \cite{Stegner99,Hansen01}. 

The purpose of the present contribution
is to analyze a simple model that was recently
introduced by Andersen and coworkers
to describe the stability and evolution of vortex ripple patterns
\cite{Andersen01a}. We focus here on the mathematical aspects of the
problem, and refer the reader to the literature 
for further motivation and a detailed
comparison to experiments \cite{Andersen01a,Andersen01b}.
The model of interest is introduced in the next section.
Sections 3 and 4 discuss the selection of the final ripple pattern
for the cases of uniform and localized perturbations of an unstable
initial state, while Section 5 contains some remarks concerning the
description of vortex ripples using continuum equations.

\section{The mass transfer model}

We consider a fully developed ripple pattern of the kind obtained in 
a quasi one-dimensional annular geometry 
\cite{Stegner99,Andersen01b,Scherer99} (see Figure 1, which is taken
from \cite{Andersen01b}).
Each of the $N$ ripples is described by a single size parameter $\lambda_n(t)$,
which can be thought to represent its length, or the amount of sand it 
contains. The periodic (annular) boundary conditions imply that
the total length $L = \sum_{n=1}^N \lambda_n$ is conserved\footnote{Models in 
which the ripple lengths and masses are
conserved separately have been developed in \cite{Andersen01a}.}. 
During one period of fluid motion, ripple $n$ exchanges mass (or length)
with its neighbors $n-1$ and $n+1$. This leads to the balance equation
\begin{equation}
\label{Model}
\frac{d \lambda_n}{dt} = 2 f(\lambda_n) - f(\lambda_{n+1}) - 
f(\lambda_{n-1}).
\end{equation}
The \emph{interaction function} $f(\lambda)$ is the central ingredient
in the model. It describes the amount of sand that is transferred 
\emph{to} a ripple of size $\lambda$ due to the vortex forming behind
this ripple.

\begin{figure}[htb]
\centerline{\psfig{file=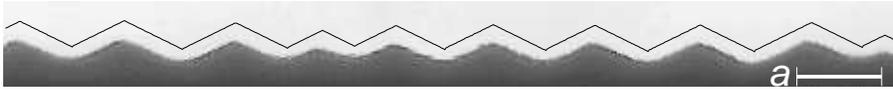,width=12.cm,angle=0}}
\vspace*{8pt}
\caption{Experimental image of a ripple pattern obtained in 
an annular container under oscillatory driving. 
The line above the sand surface shows that the pattern can
be fitted to an array of triangles with constant slope. The amplitude
of fluid motion $a$ is indicated.}
\label{ProfileFig}
\end{figure}

The following argument suggests that $f(\lambda)$ should
be a nonmonotonic function with a maximum near $\lambda = a$ 
\cite{Andersen01a}. Small ripples create a small separation vortex
which is unable to erode much of the neighboring ripples, hence
$f(\lambda)$ vanishes for small $\lambda$. On the other hand, even
for a large ripple the size of the vortex cannot be much larger
than the amplitude $a$ of the fluid motion. If $\lambda \gg a$, the
vortex does not reach beyond the trough to the next ripple, so 
$f(\lambda)$ vanishes also. The mass transfer is most efficient, and
$f$ is maximal, when $\lambda \approx a$. The interaction function can
be measured in fluid dynamical simulations \cite{Andersen01a} and
experiments \cite{Andersen01b}, which confirm these qualitative
considerations. In the following we shall take $f(\lambda)$ to be
an arbitrary, single humped function which vanishes\footnote{Note that the
dynamics (\ref{Model}) is invariant under shifts $f \to f + 
\mathrm{const}$. In general, $\lambda_{\mathrm{max}}$ is therefore
determined by the condition $f(\lambda_{\mathrm{max}}) = 
f(0)$.} at $\lambda = 0$ and at   
$\lambda = \lambda_{\mathrm{max}}$, and displays a maximum at
$\lambda = \lambda_c$. The physical meaning of $\lambda_{\mathrm{max}}$
is that new ripples are created in the troughs when the ripple spacing
exceeds $\lambda_{\mathrm{max}}$. Since we are concerned here with
the \emph{coarsening} of ripples, this process plays no r\^{o}le. 

Any homogeneous pattern with $\lambda_n \equiv \bar \lambda$ is a stationary
solution of (\ref{Model}). To investigate its stability, we 
impose a small perturbation, $\lambda_n = \bar \lambda + \epsilon_n$, and
linearize (\ref{Model}) in $\epsilon_n$. We find solutions of
the form $\epsilon_n \sim \exp[iqn + \omega(q) t]$, where the growth rate
of a perturbation of wavenumber $q$ is given by
\begin{equation}
\label{Dispersion}
\omega(q) = 2 f'(\bar \lambda) (1 - \cos q).
\end{equation}
This implies that (i) a homogeneous state is stable iff 
$f'(\bar \lambda) < 0$, and (ii) an unstable state decays predominantly
through perturbations of wavenumber $q = \pi$, in which every second 
ripple grows and every second ripple shrinks. The model (\ref{Model}) 
thus predicts an entire band of stable homogeneous states with
wavelengths $\lambda_c < \bar \lambda < \lambda_{\mathrm{max}}$, in 
qualitative accordance with experiments 
\cite{Stegner99,Hansen01,Andersen01b}.
The existence of a multitude of linearly stable, stationary states is 
a property that the model shares with many other systems in granular
physics.

When investigating the dynamics of unstable states, (\ref{Model})
has to be supplied by a rule which decides what should happen when the length
of a shrinking ripple reaches zero. We impose simply that such a ripple is 
eliminated, and the remaining ripples are relabeled so that the earlier
neighbors of the lost ripple now are next to each other. Since the time
derivatives of the lengths of these neighboring ripples jump at the instant
of disappearance of the lost ripple, this elimination procedure introduces
a distinctly non-smooth element into the dynamics. 

\section{Wavelength selection from generic unstable states}

In this section we are concerned with the evolution out of generic
unstable states. In practice this means that the initial state is of
the form $\lambda_n = \bar \lambda + \delta_n$, where $\bar \lambda < 
\lambda_c$ and the $\delta_n$ are random numbers smaller than 
$\lambda_c - \bar \lambda$. The evolution is then followed numerically
to the point where all surviving ripples are in the stable regime,
$\lambda_n > \lambda_c$. Beyond this time no further ripples are eliminated,
and hence the mean ripple length $\langle \lambda_n \rangle = L/N$
no longer changes\footnote{When all ripples are in the stable regime,
(\ref{Model}) describes a diffusive evolution towards a completely
homogeneous state.}. 

Simulations using a wide range of interaction functions
strongly indicate that the evolution selects a unique \emph{equilibrium wavelength} $\lambda_{\mathrm{eq}}$ which is independent of the initial
conditions, and depends only on the shape of the interaction function. 
In view of the large number of linearly stable stationary states, and
the deterministic character of the dynamics, this is a highly nontrivial
property of the model, which requires explanation. While a full understanding
is still lacking, we present here a partial solution which appears to yield
at least an upper bound on $\lambda_{\mathrm{eq}}$.

The key observation is that (\ref{Model}) can be cast into the form
of an overdamped mechanical system by introducing the positions of the 
ripple troughs $x_n(t)$ as basic variables, such that 
$\lambda_n = x_{n+1} - x_n$. Then the dynamics (\ref{Model}) becomes
\begin{equation}
\label{Pot}
\frac{dx_n}{dt} = - \frac{\partial V}{\partial x_n},
\end{equation}
where $V$ is a sum of repulsive pair potentials acting between the
troughs, 
\begin{equation}
\label{Pot2}
V = \sum_n v(x_{n+1}-x_n) = -  \sum_n \int_0^{x_{n+1}-x_n} d\lambda
\; f(\lambda).
\end{equation}
This is supplemented by the elimination rule, which corresponds in
the particle picture to a coalescence process\footnote{Related particle
systems have been considered in \cite{Rost95}.}.

Equation (\ref{Pot}) suggests that the dynamics is driven towards minimizing
$V$. It is then natural to surmise that the final state is determined
by the minimum of $V$ \emph{under the constraint of fixed total length
$L$}. Since the final state is clearly homogeneous, the quantity to be
minimized is
\begin{equation}
\label{Vhom}
V_{\mathrm{hom}}(\lambda) = N \int_0^\lambda d\lambda' \; f(\lambda')
= \frac{L}{\lambda} \int_0^\lambda d\lambda' \; f(\lambda').  
\end{equation}
This leads to the prediction that $\lambda_{\mathrm{eq}} = 
\lambda^\ast$, where $\lambda^\ast$ is the solution of 
\begin{equation}
\label{Lever}
\int_0^{\lambda^\ast} d\lambda \; f(\lambda) =
\lambda^\ast f(\lambda^\ast).
\end{equation}
This is of course the analytic form of the Maxwell construction, as it would
be applied
to the chemical potential in a coexistence region. 

The prediction
(\ref{Lever}) has several desirable properties. First, it is manifestly
independent of the initial condition. Second, it guarantess that
$\lambda^\ast$ is located in the stable region, i.e.
$\lambda_c < \lambda^\ast < \lambda_{\mathrm{max}}$. Third, it is
invariant under multiplication of $f$ by an arbitrary factor\footnote{This is 
required because such a multiplication only affects the time scale of
evolution and should not change the final state.}. Nevertheless it is wrong:
Comparison with simulations shows that $\lambda^\ast > \lambda_{\mathrm{eq}}$
always. Table 1 contains some typical results obtained using
a family of piecewise linear interaction functions,
\begin{equation}
\label{piece}
f(\lambda) = \left\{ \begin{array}{l@{\quad:\quad}l}
\lambda/\lambda_c  &  \lambda < \lambda_c \\ 
(\lambda_{\mathrm{max}} - \lambda)/(\lambda_{\mathrm{max}}-
\lambda_c)  & \lambda \geq \lambda_c. \end{array} \right.
\end{equation} 
In this case
\begin{equation}
\label{pieceast} 
\lambda^\ast/\lambda_c = 
\sqrt{\lambda_{\mathrm{max}}/\lambda_c}.
\end{equation}
The two wavelengths $\lambda^\ast$ and 
$\lambda_{\mathrm{eq}}$ appear to become equal, in the sense that
$(\lambda^\ast - \lambda_c)/(\lambda_{\mathrm{eq}} - \lambda_c) \to 1$,
when the stable branch of the transfer function becomes very steep, i.e.
when $\lambda_{\mathrm{max}} - \lambda_c \ll \lambda_c$. On the other
hand, when the stable branch is shallow, the prediction 
$\lambda_{\mathrm{eq}} = \lambda^\ast$ fails completely. This can be
seen by considering the extreme case of an interaction function that remains
constant for $\lambda > \lambda_c$, i.e., (\ref{piece}) with
$\lambda_{\mathrm{max}}/\lambda_c = \infty$. Then we find numerically that
the final wavelength is $\lambda_{\mathrm{eq}} \approx 1.61 \; \lambda_c$,
while clearly $\lambda^\ast = \infty$.

\begin{table}[htb]
\caption{Comparison of the wavelength $\lambda^\ast$ predicted by 
(\ref{Lever}) with the equilibrium wavelength
$\lambda_{\mathrm{eq}}$ obtained in numerical simulations of the model.
The simulation results were averaged over 100 runs
using 1000 initial ripples with lengths uniformly distributed in 
[0.5,1].}
{\begin{tabular}{|c||c|c|c|c|c|c|c|c|}
\hline
$\lambda_{\mathrm{max}}/\lambda_c$ & 1.1  & 1.25 & 1.5 & 2 & 4 & 6&  8 & $\infty$ \\ 
\hline 
$\lambda^\ast/\lambda_c $ & 1.0488 & 1.118 & 1.225 & 1.414 & 2 & 2.449 &  2.828 &
$\infty$ \\
\hline
$\lambda_{\mathrm{eq}}/\lambda_c $ & 1.0477 & 1.1105 & 1.196 & 1.307 & 1.461 &
1.516 &  1.536 & 1.607 \\
\hline
\end{tabular}}
\end{table}

An obvious interpretation of the finding $\lambda_{\mathrm{eq}} < 
\lambda^\ast$ is that the deterministic, 
overdamped dynamics (\ref{Pot}) gets stuck
in a metastable state before reaching the configuration of minimal 
``energy'' $V$. This suggests that it should be possible to increase
the final wavelength by either making the dynamics less damped, or by
introducing noise. The first modification implies that the 
trough ``particles''
are supplied with a mass and a momentum variable along the lines of 
\cite{Rost95}. Preliminary simulations show that this does indeed increase
the final wavelength, but not sufficiently to reach $\lambda^\ast$. 
A noisy version of the model has been described in \cite{Andersen01b}.
Noise also increases the final wavelength, however in addition it introduces
a new coarsening mechanism involving rare fluctuations \cite{Werner93}, 
which in principle
drives the wavelength towards $\lambda_{\mathrm{max}}$,
as long as ripple creation is not included.

\section{Front Propagation}

We now consider a perfectly ordered, homogeneous, unstable
initial condition,
$\lambda_n \equiv \lambda^{(i)} < \lambda_c$ for all $n$, which is destabilized
by a local perturbation, e.g. by making a single ripple shorter or longer.
Then two fronts emanate from the perturbed region which propagate into 
the unstable state and leave in their wake a stable homogeneous configuration
at a new wavelength $\lambda^{(f)} > \lambda_c$
(Figure 2). The elimination of ripples
occurs in the vicinity of the fronts. 

\begin{figure}[htb]
\label{Front1}
\centerline{\psfig{file=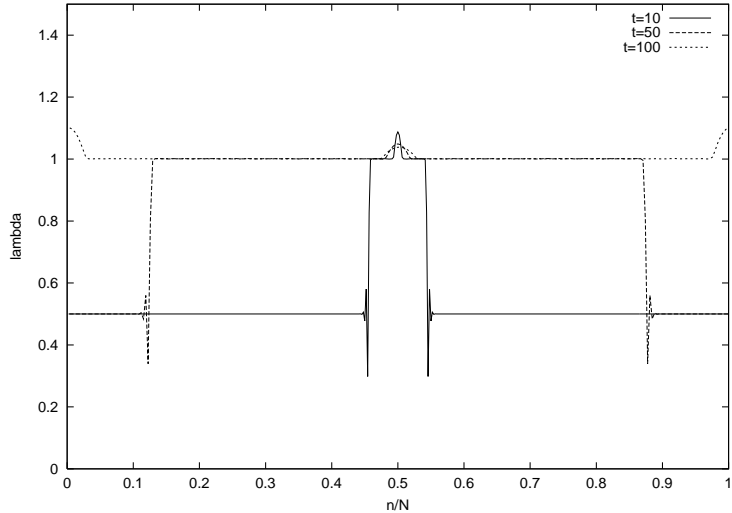,width=7.cm,angle=-90}}
\vspace*{8pt}
\caption{Front propagation for a parabolic interaction function
$f(\lambda) = 2 \lambda - \lambda^2$ with initial wavelength
$\lambda^{(i)} = 1/2$ and final wavelength $\lambda^{(f)} = 1$. 
The figure shows the ripple wavelength $\lambda_n$ as a function of the
scaled ripple number $n/N$ (note that $N$ decreases with time).}
\end{figure}

Since the period-2 mode ($q = \pi$) is the most unstable according
to (\ref{Dispersion}), its growth controls the propagation of the fronts.
Following the standard theory of front propagation into unstable states
\cite{vanSaarloos88}, we write the propagating perturbation
as a traveling wave with an exponential tail,
\begin{equation}
\label{Fronts}
\epsilon_n(t) = (-1)^n \exp[-\alpha(n - ct)],
\end{equation}
where $c$ is the propagation speed and $\alpha$ the decay constant.
Inserting this into the linearization of (\ref{Model}) we find
the relation
\begin{equation}
\label{speed}
c(\alpha) = \frac{2 f'(\lambda^{(i)})}{\alpha} (1 + 
\cosh(\alpha))
\end{equation}
between $c$ and $\alpha$. Localized initial conditions usually propagate
at the ``marginal stability'' speed $c^\ast$ corresponding to the minimum
of (\ref{speed}) \cite{vanSaarloos88}, and hence we expect
that the front velocity is given by $c^\ast \approx 4.4668 \,
 f'(\lambda^{(i)})$.
This prediction is well confirmed by numerical simulations.

We next turn to the relationship between the initial and final wavelengths.
In stark contrast to the situation discussed in Section 3, here 
the final selected wavelength depends on the initial state in a surprisingly
complex manner (Figure 3). The most prominent feature in the graph is 
a straight line of slope 2 which extends from $\lambda_c/2 = 0.5$ to 
$\lambda_{\mathrm{max}}/2 = 0.75$. In this regime the wavelength selection
process is very simple. The growth of the period-2 mode near the front implies
that every second ripple is eliminated, hence the wavelength doubles. 
When $\lambda^{(i)} > \lambda_c/2$, this is sufficient to bring the
ripples into the stable band. We therefore conclude that
\begin{equation}
\label{ratio1}
\lambda^{(f)} = 2 \lambda^{(i)} \;\;\;\;
\mathrm{for} \;\;\;\; 
\lambda_c/2 < \lambda^{(i)} < \lambda_{\mathrm{max}}/2.
\end{equation}

\begin{figure}[htb]
\centerline{\psfig{file=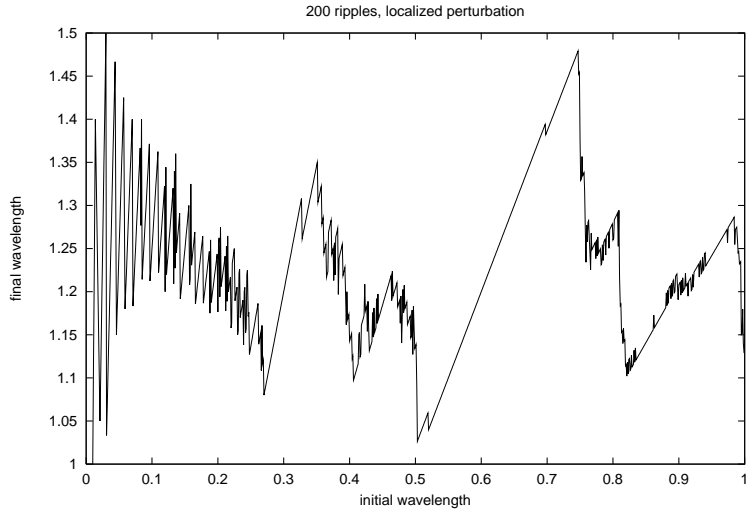,width=7.cm,angle=-90}}
\vspace*{8pt}
\caption{Final wavelength as a function of initial wavelength for the 
piecewise linear interaction function (\ref{piece}) 
with $\lambda_{\mathrm{max}}/
\lambda_c = 3/2$. For each value of $\lambda^{(i)}$ a system of initially
200 ripples was simulated until all ripples reached the stable regime.
All wavelengths are measured in units of $\lambda_c$}
\end{figure}

When $\lambda^{(i)} < \lambda_c/2$, the ripples are still unstable after
the doubling of the wavelength. The simplest scenario for the further
evolution is that the new state again becomes unstable with respect
to the period-2 mode, so that the wavelength doubles once more. Indeed
a segment with slope 4 can be detected in Figure 3, which starts near
$\lambda_c/4$. For smaller initial wavelengths this scenario breaks down
because the accumulation of exponentially growing perturbations prevents
the intermediate homogeneous states to become established. We have not 
attempted any further analysis of the complicated
behavior seen in Figure 3 for $\lambda^{(i)} < \lambda_c/4$.

A different kind of complication arises when $\lambda^{(i)} > 
\lambda_{\mathrm{max}}/2$. In this case the growth of the period-2 mode
terminates before the smaller ripples have reached zero length, because
the system gets temporarily trapped in a \emph{stationary} period-2
state with alternating ripple lengths $\lambda_A > \lambda_c$ and
$\lambda_B < \lambda_c$ (Figure 4). It is possible to prove that such a state,
which has to satisfy the constraints
\begin{equation}
\label{period2}
f(\lambda_A) = f(\lambda_B) , \;\;\;
(\lambda_A + \lambda_B)/2 = \lambda^{(i)}
\end{equation}
always exists when $\lambda_{\mathrm{max}} < 2 \lambda_c$ and 
$\lambda^{(i)} > \lambda_{\mathrm{max}}/2$. 
Indeed, consider the function 
\begin{equation}
\label{F}
F(\lambda_A) = f(\lambda_A) - f(\lambda_B) = 
f(\lambda_A) - f(2 \lambda^{(i)} - \lambda_A).
\end{equation}
This function vanishes at $\lambda_A = \lambda_B = \lambda^{(i)}$, where its
slope $F'(\lambda^{(i)}) = 2 f'(\lambda^{(i)}) > 0$. Furthermore $F$ is
odd under reflection around $\lambda_A = \lambda^{(i)}$, and
$F(\lambda_{\mathrm{max}}) = - f(2 \lambda^{(i)} - \lambda_{\mathrm{max}}) 
< 0$, because $2 \lambda^{(i)} - \lambda_{\mathrm{max}} > 0$. It then
follows by continuity that $F$ has to possess two additional zeros,
corresponding to a solution of (\ref{period2}) with 
$\lambda_A > \lambda_B$. It also follows that 
\begin{equation}
\label{stab2}
f'(\lambda_A) + f'(\lambda_B) < 0
\end{equation}
which is the condition for stability \emph{within} the space of period-2
configurations.

\begin{figure}[htb]
\centerline{\psfig{file=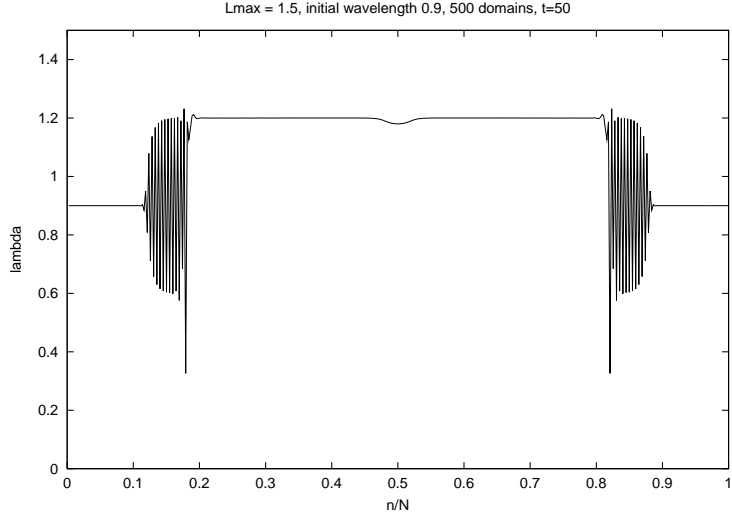,width=7.cm,angle=-90}}
\vspace*{8pt}
\caption{Front propagation for $\lambda^{(i)} > \lambda_{\mathrm{max}}/2$.
The figure shows a system of initial 500 ripples at time $t=50$. The 
interaction function was piecewise linear with $\lambda_{\mathrm{max}}/
\lambda_c = 3/2$. Note the period-2 state appearing between the front and
the homogeneous final state. Here $\lambda^{(i)} = 0.9 \, \lambda_c$ and
$\lambda^{(f)} = 1.2 \, \lambda_c = 4/3 \, \lambda^{(i)}$. }
\end{figure}

A stability analysis of the
stationary period-2 state 
yields the linear growth rate
\begin{equation}
\label{Dispersion2}
\omega(q) = f'(\lambda_A) + f'(\lambda_B) + 
\sqrt{(f'(\lambda_A) - f'(\lambda_B))^2 + 4
f'(\lambda_A)f'(\lambda_B) \cos^2 q}.
\end{equation}
Since $f'(\lambda_A) f'(\lambda_B) < 0$, the growth rate is
maximal at $q = \pi/2$, and it vanishes at $q = 0$ and $q = \pi$.
We conclude that the stationary period-2 solution 
is most unstable with respect to 
perturbations of period 4. In effect, this implies that one out of four ripples
is eliminated, and hence $\lambda^{(f)}/\lambda^{(i)} = 4/3$. 
This explains the region of slope 4/3 in Figure 3 starting
around $\lambda^{(i)} \approx 0.8$. 
Other rational
ratios can (and do) appear in a similar manner.

\section{Continuum equations for vortex ripples?} 

The model (\ref{Model}) was proposed to describe the stability and
evolution of fully developed ripple patterns, but it does not address
the question of how these patterns emerge from the flat bed. In part, 
this reflects the fact that the separation vortices
appear only once the pattern has reached a certain 
amplitude, so a different mechanism must control the
initial instability \cite{Blondeaux90}. On the other hand,
a theoretical description that encompasses the transient
evolution from the flat bed \emph{as well as}
the fully developed ripple pattern would be highly desirable, in 
particular for the analysis of two-dimensional systems \cite{Hansen01}. 
In this section we suggest that such a comprehensive description may
be difficult to achieve.

For the related problem of wind-driven (\emph{aeolian}) sand ripples, 
a description in terms of partial differential equations for the 
(one-dimensional) continuous profile $h(x,t)$ of the sand surface has
been developed \cite{Csahok00}. Let us collect the properties that such
an equation should have for the case of vortex ripples under water. 
(i) Since the pattern does not depend on the thickness of the water layer, the
dynamics should be invariant under constant shifts of the height, 
$h \to h + {\mathrm{const}}$. (ii) The oscillatory driving implies symmetry
under $x \to -x$. (iii) The slope of the ripples should saturate around
the angle of repose, and (iv) the pattern should \emph{not} be invariant
under $h \to -h$ (closer inspection of profiles like that in Figure 1 show
that the peaks are cusp-like while the troughs are rounded, 
see \cite{Stegner99}). Restricting ourselves to terms which are polynomial
in the derivatives of $h$, the simplest 
equation satisfying these requirements is 
\begin{equation}
\label{cont}
h_t = - h_{xx} - h_{xxxx} + (h_x)^3_x - b (h_x)^2_{xx},
\end{equation}
where subscripts refer to partial derivatives and $b$ is a positive 
constant. It is easy to see that the flat bed solution of (\ref{cont}) 
is unstable, with the fastest growing mode (corresponding to the initial
pattern) occurring at wavelength $2\pi \sqrt{2}$. The third term on the right
hand side leads to a selected slope of $\pm 1$, while the last term
sharpens the peaks and rounds off the troughs of the ripples. 

A detailed study of (\ref{cont}) has been carried out by Politi \cite{Politi98},
who shows that the wavelength of the pattern coarsens indefinitely as 
$\ln t$. This conclusion appears to apply generally to equations of the same
general form with polynomial terms \cite{Csahok00}. 
Patterns which do not coarsen can
be obtained only at the expense of introducing unbounded growth of the 
slope, and hence of the amplitude, of the pattern \cite{Krug99}. 
A class of equations
which contains both types of behavior is 
\begin{equation}
\label{stepeq}
h_t =  -\left\{ \frac{ h_x}{1+h_x^2}  +
\frac{1}{(1+h_x^2)^\nu}
\left[ \frac{h_{xx}}{(1+h_x^2)^{3/2}}
\right]_x \right\}_x,
\end{equation}
which arises in the context of meandering instabilities of stepped
crystal surfaces \cite{Kallunki00}. The exponent $\nu$ is characteristic
of the relaxation mechanism of the steps, the cases of immediate
physical relevance corresponding to 
$\nu = 1$ and $\nu = 1/2$ \cite{Kallunki00}. The analysis of this equation
shows that unbounded amplitude growth occurs 
for $-1/2 < \nu < 3/2$, and coarsening for $\nu < -1/2$. 

We therefore conjecture that local height equations generally cannot describe
the emergence and evolution
of patterns of constant wavelength \emph{and} amplitude.
A general proof, or the discovery of a counterexample, would be of 
considerable interest. Meanwhile, we believe that models 
like (\ref{Model}) can play a useful part in the analysis 
of such patterns.

% \begin{figure}[b]
% \centerline{\psfig{file=run.ps,width=8.cm,angle=-90}}
% \vspace*{8pt}
% \caption{}
% \end{figure}

% \begin{figure}[b]
% \centerline{\psfig{file=lambda.ps,width=8.cm,angle=-90}}
% \vspace*{8pt}
% \caption{}
% \end{figure}

% \begin{figure}[b]
% \centerline{\psfig{file=convergence.ps,width=8.cm,angle=-90}}
% \vspace*{8pt}
% \caption{}
% \end{figure}

\section*{Acknowledgements}

I am much indebted to Ken H. Andersen and Tomas Bohr for many 
enlightening discussions and interactions. Most of this work was
performed during a sabbatical stay at CAMP, Denmarks Technical University,
Lyngby, and at the Niels Bohr Institute, Copenhagen. The kind and generous
hospitality of these institutions is gratefully acknowledged.

\end{document}